\documentclass[preprint,aps,floatfix]{revtex4}
\usepackage{graphicx}
\usepackage{graphics}
\usepackage{color}
\usepackage{subfigure}
\usepackage{amsmath}
\usepackage{amssymb, amsthm}
\usepackage{txfonts}

\begin{document}

\title{A Dynamical Self-Consistent Finite Temperature Kinetic Theory: \\ The ZNG Scheme \label{Chapter_Allen_Barenghi}}
\thanks{This chapter is dedicated to Allan Griffin, an inspirational colleague, mentor and
friend who was actively working\\\vspace{-0.5cm} on these issues until he passed away on 19th May 2011.}
\author{A. J. Allen$^1$, C. F. Barenghi$^1$, N. P. Proukakis$^1$, and E. Zaremba$^2$}
\address{$1$ Joint Quantum Centre (JQC) Durham-Newcastle, School of Mathematics and Statistics, Newcastle University, Newcastle Upon Tyne, NE7
1RU, UK.\\  $2$ Department of Physics, Engineering Physics and Astronomy, Queens University, Kingston, Ontario, K7L 3N6, Canada.}

\begin{abstract}
We review a self-consistent scheme for modelling trapped \index{interaction(s)!weak}weakly-interacting quantum gases at temperatures where the condensate coexists with a significant \index{thermal!cloud}thermal cloud.  This method has been applied to atomic gases by \index{Zaremba--Nikuni--Griffin|see{ZNG}}Zaremba, Nikuni, and Griffin, and is often referred to as \index{ZNG}ZNG. 
It describes both mean-field-dominated and \index{hydrodynamic(s)!regime}hydrodynamic regimes, except at very low temperatures or in the regime of large \index{fluctuations!thermal}\index{fluctuations!quantum}fluctuations.
Condensate dynamics are described by a \index{Gross--Pitaevskii equation! dissipative}dissipative Gross--Pitaevskii equation (or
the corresponding quantum \index{hydrodynamic(s)!equation}hydrodynamic equation with a \index{source term}source term), while the \index{non-condensate!dynamics}non-condensate
evolution is represented by a \index{Boltzmann equation! quantum|see{quantum, Boltzmann equation}}\index{quantum! Boltzmann equation}quantum Boltzmann equation, which additionally
includes \index{collision(s)}collisional processes which  transfer atoms between these two subsystems. In the \index{mean-field dominated regime|see{collisionless regime}}\index{collisionless regime}mean-field-dominated regime collisions are treated perturbatively and the full distribution function is needed to describe the \index{thermal!cloud}thermal cloud, while in the \index{hydrodynamic(s)!regime}hydrodynamic regime the system is parametrised in terms of a set of local variables.
Applications to finite
temperature induced \index{damping}\index{Landau!damping}damping of collective modes and \index{vortex!finite temperature dynamics}vortices in the
mean-field-dominated regime are  presented.\\~\\
Unedited version of chapter to appear in\\
Quantum Gases: Finite Temperature and Non-Equilibrium Dynamics (Vol. 1 Cold Atoms Series).\\
N.P. Proukakis, S.A. Gardiner, M.J. Davis and M.H. Szymanska, eds.\\ Imperial College Press, London (in press).\\
See http://www.icpress.co.uk/physics/p817.html
\vspace{-5cm}
\end{abstract}
\maketitle

\section{Introduction}
\label{allen_sec:intro}
In most experiments involving Bose--Einstein condensates, the system is only partially condensed, meaning that thermal excitations can play an important role in the \index{damping}damping of condensate collective
modes~\cite{jin_matthews_97,stamper-kurn_miesner_98,marago_hechenblaikner_01,chevy_bretin_02}
or macroscopic excitations such as
\index{soliton!dark}\index{soliton!bright}solitons~\cite{burger_bongs_99}
and \index{vortex}vortices~\cite{rosenbusch_bretin_02,abo-shaeer_raman_02}, making it
imperative to include the full dynamics of the non-condensed atoms when
modelling such systems.  
This includes,
beyond the usual  mean-field contributions,
\index{collision(s)}collisions between the \index{collision(s)!thermal-thermal}\index{non-condensate!--non-condensate interaction|see{collision(s)!thermal-thermal}}non-condensate atoms, and 
 particle-exchanging
\index{collision(s)}collisions between the condensate and \index{non-condensate!--condensate particle transfer|see{collision(s)!condensate-thermal particle exchanging}}non-condensate\index{collision(s)!condensate-thermal}\index{collision(s)!condensate-thermal particle exchanging}.
The approach reviewed in this chapter reflects an accurate representation of these combined dynamics, and can therefore fully 
simulate the back-action of the \index{thermal!cloud}thermal cloud on the condensate, which is often neglected, treated to lowest order or in \index{linear response}linear response (see 
Ref.\  \cite{proukakis_jackson_08}).
As this method has been implemented by \index{ZNG}Zaremba, Nikuni and Griffin (following
on from early work by \index{Kirkpatrick}Kirkpatrick and \index{Dorfman}Dorfman
\cite{kirkpatrick_dorfman_83,kirkpatrick_dorfman_85a,kirkpatrick_dorfman_85b,kirkpatrick_dorfman_85c})
we henceforth refer to it as the \index{ZNG}ZNG 
method, as described in detail in the
recent book of these authors \cite{griffin_nikuni_book_09}. 

The strength and frequency of \index{collision(s)}collisions between atoms
characterises two distinct dynamical regimes \cite{griffin_book_93}:
(i) in the \index{collisionless regime}collisionless (or mean-field dominated) regime, in which most
experiments with ultracold atomic gases are conducted (in stark
contrast to helium), the physics tends to be dominated by mean-field effects;
nonetheless, an accurate description of \index{collision(s)}collisions is {\it essential} for
fully describing the system properties, and accounting for changes in condensate atom number. Importantly,  
a clear \index{collision(s)!separation of timescales}separation of timescales (average collision time is longer that the collective mode period), enables collisions in this regime to be treated {\it perturbatively} (see also Ref.\  \cite{proukakis_jackson_08}).
(ii) Some recent experiments have been conducted in the crossover to
\cite{stamper-kurn_miesner_98} or deep within the \index{hydrodynamic(s)!regime}hydrodynamic (or collision
dominated) \cite{meppelink_koller_09} regime.
In this regime --- which has strong analogies to the \index{two-fluid hydrodynamic model}two-fluid behaviour of \index{superfluid!helium}superfluid $^4$He ---, the high gas density leads to very rapid \index{collision(s)}collisions between thermal atoms, such that the \index{non-condensate}non-condensate enters a {\it local} \index{hydrodynamic(s)!equilibrium}hydrodynamic equilibrium (a precursor to true \index{therodynamic equilibrium|see{equilibrium, thermodynamic}}\index{equilibrium! thermodynamic}thermodynamic equilibrium); this enables its description in terms of a few local \index{hydrodynamic(s)!variables}hydrodynamic variables (e.g.\ local density, velocity, \index{chemical potential}chemical potential, temperature, and pressure).

The power of the \index{ZNG}ZNG theory lies in its ability (i) to successfully describe both \index{collisionless regime}collisionless experiments with collective modes
\cite{williams_griffin_01,jackson_zaremba_01,jackson_zaremba_02b} and
\index{soliton!dark}\index{vortex}macroscopic excitations \cite{jackson_proukakis_07,jackson_proukakis_09} at
finite temperatures (see Section~\ref{allen_applications}); and (ii) to reduce to
the damped \index{two-fluid hydrodynamic model}two-fluid equations of $^4$He in the \index{hydrodynamic(s)!regime}hydrodynamic limit
\cite{zaremba_nikuni_99,nikuni_zaremba_99,nikuni_griffin_01a,nikuni_griffin_01b,nikuni_griffin_98a,nikuni_02,griffin_zaremba_97,nikuni_griffin_04,zaremba_griffin_98}
(Section~\ref{allen_relevance}).

\section{Methodology}
\label{allen_sec:methodology}
The \index{second quantisation}second quantised \index{Hamiltonian, expression for!second-quantised (general)}Hamiltonian for a  \index{interaction(s)!weak}weakly-interacting Bose gas is given by
\begin{equation}
\hat {H}  =  \int d {\mathbf{r}} \hat {\Psi}^\dagger (\mathbf{r}) 
\left\{
\left[ - \frac{\hbar^2 \nabla ^2}{2m} +
V_{\rm {ext}}(\mathbf{r})\right]
+ \frac{1}{2} \int d {\mathbf{r}}' \hat{\Psi}^\dagger (\mathbf{r}') U(\mathbf{r},\mathbf{r}') \hat {\Psi}(\mathbf{r}')
\right\}
\hat {\Psi}(\mathbf{r}),
\label{allen_eq:hamiltonian}
\end{equation}
where $V_{\rm {ext}}(\mathbf{r})$ is the external confining potential
and
$U(\mathbf{r},\mathbf{r}') = g \delta({\mathbf{r}} - {\mathbf{r}}')$ describes binary \index{interaction(s)!binary}interactions, with $g = 4\pi \hbar^2a/m$, where $a$ is the $s$-wave \index{scattering!$s$-wave}scattering length.
Following \index{Beliaev}Beliaev \cite{beliaev_58}, we decompose the \index{second quantisation}second quantised \index{field operator!symmetry-breaking decomposition}field operator as
\begin{equation}
\hat {\Psi}(\mathbf{r},t) =  \phi (\mathbf{r},t) + \hat \psi ' (\mathbf{r},t),
\label{allen_eqn:decomp}
\end{equation}
where we define a \index{condensate!wavefunction}condensate wavefunction via the non-equilibrium \index{average!ensemble|see{ensemble, average}}\index{ensemble! average}ensemble average $
\phi(\mathbf{r},t) = \langle \hat {\Psi}(\mathbf{r},t)\rangle $, which takes a non-zero value
under the assumption of Bose broken symmetry\index{symmetry-breaking}.  This in turn implies that $\langle \hat \psi ' (\mathbf{r},t) \rangle = 0$, 
with $\hat \psi'$ capturing all \index{fluctuations!thermal}fluctuations around
the classical mean-field of the condensate, thus is often termed the
\index{field operator!non-condensate}non-condensate operator.

By taking an average in the Heisenberg equation of motion for the
\index{field operator}field operator, $i\hbar d \hat\Psi/ d t = [\hat \Psi,\hat H]$ one obtains the equation of motion for the macroscopic wavefunction (omitting the explicit dependence on ${\mathbf{r}}$ and $t$), 
\begin{equation}
i \hbar \frac {\partial \phi}{\partial t} = \left(-\frac{\hbar ^2 \nabla ^2}{2m} +
V_{\mathrm{ext}}\right) \phi 
+ g \langle \hat \Psi^\dagger\hat \Psi \hat \Psi \rangle,
\end{equation}
which is {\it exact} in the context of \index{symmetry-breaking}symmetry-breaking.
Expanding the final term using~(\ref{allen_eqn:decomp}), we obtain
$ \langle \hat \Psi^\dagger \hat \Psi \hat \Psi \rangle
  = n_{\mathrm{c}} \phi +  m' \phi^* + 2 n' \phi + \langle \hat \psi'^\dagger \hat \psi' \hat \psi' \rangle$,
where we have defined the
following mean-field `densities'\index{non-condensate!density}\index{anomalous pair density|see{anomalous average, pair}}\index{anomalous average!pair}: 
\vspace{-0.2cm}
\begin{itemize}
\item $\mbox{condensate density:}  \hspace{1.6cm} n_{\mathrm{c}}(\mathbf{r},t) = |\phi(\mathbf{r},t) |^2$,
 \item $\mbox{non-condensate density:}  \hspace{1.0cm} n'(\mathbf{r},t) =\langle  \hat \psi'^\dagger (\mathbf{r},t) \hat \psi'(\mathbf{r},t)\rangle$, 
\item $\mbox{anomalous pair density:} \hspace{1.05cm} m'(\mathbf{r},t) =  \langle \hat \psi' (\mathbf{r},t) \hat \psi' (\mathbf{r},t)\rangle$.
\end{itemize}
\vspace{-0.2cm}
 The equation of motion for the condensate therefore reduces
to~\cite{proukakis_burnett_96,griffin_96,proukakis_thesis_97,proukakis_burnett_98}
 \begin{equation}
 i \hbar \frac {\partial \phi}{\partial t} = \left [ -\frac{\hbar ^2 \nabla ^2}{2m} +
V_{\mathrm{ext}} + g(n_{\mathrm{c}}  + 2 n')  \right]\phi +  g m'\phi^* + g \langle \hat \psi'^\dagger \hat \psi' \hat \psi'\rangle. 
 \label{allen_eqn:condeom}
 \end{equation}
 Thus the condensate is coupled to higher order correlations which are
further coupled to even higher order non-condensate\index{non-condensate!higher-order correlations} correlations, {\it ad infinitum}, suggesting the need for a suitable truncation. 
An important feature of this equation, highlighted in
\cite{proukakis_burnett_96,proukakis_thesis_97,proukakis_burnett_98},
is that condensate growth from a zero initial value can only arise through the
conventionally-neglected `anomalous triplet term' $\langle \hat \psi'^\dagger \hat \psi' \hat \psi'\rangle$, pointing to its importance as a `\index{source term}source term' in condensate \index{condensate!kinetics}\index{kinetics!of condensate|see{condensate, kinetics}}kinetics \cite{zaremba_nikuni_99}.
The corresponding equation of motion for \index{non-condensate!atom}non-condensate atoms reads 
\begin{equation}
\begin{split}
i\hbar \frac{\partial \hat \psi' }{\partial t} =
& \left [ - \frac{ \hbar ^2 \nabla
^2}{2m} + V_{\rm {ext}} + 2g(n_{\mathrm{c}} + n')\right]  \hat \psi' - 2gn' \hat \psi' + g \phi^2 \hat \psi'
\\ 
&
+ g \phi^* \left(\hat \psi' \hat \psi' - m'\right)
+ 2 g \phi \left(\hat \psi'^\dagger \hat \psi' - n'\right)
+ g \left(\hat \psi'^\dagger \hat \psi' \hat \psi' - \langle \hat
\psi'^\dagger \hat \psi'\hat \psi'\rangle \right).
\label{allen_eqn:noncondeom} 
\end{split}
\end{equation}
Note that if all the terms $n'(\mathbf{r},t), \, m'(\mathbf{r},t)$, and $\langle \hat
\psi'^\dagger \hat \psi' \hat \psi'\rangle$ are set to zero, Eq.~(\ref{allen_eqn:condeom}) trivally reduces to the zero-temperature
Gross--Pitaevskii Equation (\index{Gross--Pitaevskii equation}\index{GPE|see{Gross--Pitaevskii equation}}GPE).  For interacting
systems and/or at finite temperatures
however, these quantities will in general be present and should be appropriately dealt with.

\subsection{Mean-Field Coupling: First Order Effects}
\label{allen_sec:firstorder}
The simplest $T>0$ \index{mean-field theory! Hartree--Fock}Hartree--Fock
mean-field method~\cite{giorgini_pitaevskii_97,pethick_smith_book_02,pitaevskii_stringari_book_03} is a limiting case of the above equations, in which only normal non-condensate terms involving one creation and one annihilation operator are
retained, with anomalous contributions neglected in both Eq.\ (\ref{allen_eqn:condeom}) and the entire second line of Eq.\ (\ref{allen_eqn:noncondeom}).  
In this limit, and upon making a \index{semiclassical approach}semiclassical approximation for the kinetic energy, the local energy of the thermal atoms becomes in the \index{mean-field theory! Hartree--Fock}Hartree--Fock approximation
\begin{equation}
\tilde \varepsilon_i({\mathbf{r}},t) = \frac{p^2}{2m} + V_{\rm {ext}}(\mathbf{r}) +
2g[n_{\mathrm{c}}({\mathbf{r}},t) + n'({\mathbf{r}},t)] \equiv \frac{p^2}{2m} + U_{\rm{eff}}({\mathbf{r}},t) ,
\label{allen_eqn:hfenergies}
\end{equation}
beyond kinetic energy and external potential, this includes a mean-field potential $U_{\rm{eff}}({\mathbf{r}},t) =  V_{\rm {ext}}(\mathbf{r}) +2g[n_{\mathrm{c}}({\mathbf{r}},t) +
n'({\mathbf{r}},t)] $ created by the condensate
$n_{\mathrm{c}}({\mathbf{r}},t)$ and non-condensate\index{non-condensate!density} $n'({\mathbf{r}},t)$ densities which acts on a thermal atom as it propagates.

In the \index{HFB|see{mean-field theory, Hartree--Fock--Bogoliubov}}\index{Hartree--Fock--Bogoliubov|see{mean-field theory, Hartree--Fock--Bogoliubov}}\index{mean-field theory! Hartree--Fock--Bogoliubov}Hartree--Fock--Bogoliubov (HFB) extension, all {\it quadratic} \index{non-condensate}non-condensate operators are maintained in the Hamiltonian\index{Hamiltonian!Hartree--Fock--Bogoliubov}, i.e.\ \index{mean-field theory! Hartree--Fock--Bogoliubov}HFB additionally includes anomalous terms with two like creation or annihilation operators; 
such an approach, which can also be derived variationally
\cite{blaizot_ripka_book_86}, is appealing as it relies on a {\it quadratic}
Hamiltonian\index{Hamiltonian!quadratic}, which can be routinely diagonalised by a \index{Bogoliubov!transformation}\index{transformation!Bogoliubov|see{Bogoliubov, transformation}}Bogoliubov transformation
to a \index{quasiparticle!basis}quasiparticle basis \cite{pethick_smith_book_02}. 
Despite explicitly accounting for pair \index{anomalous average!pair}anomalous averages, and providing a lower total energy for the system, this approach is problematic as its homogeneous 
 limit leads to a gap in the energy spectrum at low momenta, which
violates the \index{Goldstone! theorem}Goldstone
theorem~\cite{binney_book_92}.  This
inconsistency can be avoided by neglecting the \index{anomalous average!pair}anomalous average altogether (or
using other tricks
\cite{bijlsma_stoof_97,proukakis_morgan_98,yukalov_kleinert_06,tommasini_depassos_05,kita_06}), as discussed by
Griffin~\cite{griffin_96,shi_griffin_98} and implemented numerically in
Refs.~\cite{hutchinson_zaremba_97,hutchinson_burnett_00}. The latter simplified approximation
(despite not accounting for \index{many-body!effects}many-body corrections to the calculations \cite{burnett_chapter_99,shi_griffin_98,bijlsma_stoof_97})
 forms a good basis for finite temperature perturbative theories, with higher order effects in $g$ 
essential to account for collisional processes.

\subsection{Particle-Exchanging \index{collision(s)}Collisions: Second Order Effects}
\label{allen_sec:secondorder}
The important \index{collision(s)}collisional processes describing the transfer of an atom into / out of the condensate\index{collision(s)!condensate-thermal particle exchanging}
is contained in the triplet correlation $\langle \hat \psi'^\dagger \hat \psi' \hat \psi'\rangle$ of Eq.\
(\ref{allen_eqn:condeom})~\cite{proukakis_burnett_96,proukakis_burnett_98,zaremba_nikuni_99}. 
Careful consideration (see also \cite{proukakis_thesis_97,proukakis_01}) leads to a \index{Gross--Pitaevskii equation! dissipative}dissipative Gross--Pitaevskii equation for the \index{condensate!wavefunction}condensate wavefunction $\phi (\mathbf{r},t)$ given by \cite{zaremba_nikuni_99}
\begin{equation}
i \hbar \frac {\partial \phi({\mathbf{r}},t)}{\partial t} = \left\{ -\frac{\hbar ^2 \nabla
^2}{2m} + V_{\mathrm{ext}}({\mathbf{r}}) + g \left[n_{\mathrm{c}}(\mathbf{r},t)  + 2n'(\mathbf{r},t) \right]- iR(\mathbf{r},t) \right\}\phi({\mathbf{r}},t) , 
\label{allen_eqn:dgpe}
\end{equation}
where 
$R(\mathbf{r},t) = - (g/n_{\mathrm{c}}) 
{\rm{Im}}
(\phi^* \langle \psi'^\dagger \psi'\psi'\rangle)$, with 
${\rm{Im}}(\ldots)$ 
denoting the imaginary part,
is a non-Hermitian \index{source term}source term which allows the normalisation of the \index{condensate!wavefunction}condensate wavefunction $\phi(\mathbf{r},t)$ to change with time.

The derivation of the \index{kinetic equation}kinetic equation for the \index{non-condensate!atom} non-condensate atoms, based on Eq.\ (\ref{allen_eqn:noncondeom}), follows from the pioneering work of \index{Kirkpatrick}Kirkpatrick and
\index{Dorfman}Dorfman~\cite{kirkpatrick_dorfman_85b,kirkpatrick_dorfman_85a} on uniform Bose gases as discussed in detail 
in~\cite{zaremba_nikuni_99,griffin_nikuni_book_09}; here we summarise the main results.
Key to this is to impose a local \index{semiclassical approach}semiclassical approximation and to describe the \index{thermal!cloud}thermal cloud by a distribution function. We thus introduce the \index{Wigner!distribution function}Wigner distribution function $f({\bf{p}},{\bf{r}}, t)$ for an atom of momentum ${\bf{p}}$, at location ${\bf{r}}$, and time $t$, 
defined in terms of the 
Wigner operator\index{Wigner!operator}, $\hat f$, for the \index{non-condensate!atom}non-condensate atoms,
$ 
\hat f({\bf{p}},{\bf{r}}, t)  = \int d {\bf{r}}' e^{i {\bf{p}}\cdot
{\bf{r}}'/\hbar} \hat \psi'^\dagger \left(  {\bf{r}} + {\bf{r'}}/2, t\right)\hat \psi' \left(  {\bf{r}} -{\bf{r'}}/2, t\right),
$ 
via the expression 
$f({\bf{p}},{\bf{r}}, t) = \langle \hat f({\bf{p}},{\bf{r}}, t)\rangle = {\rm {Tr}} \{\tilde \rho (t,t_0) \hat f({\bf{p}},{\bf{r}}, t)\}$;
here
${\rm {Tr}}$ denotes the trace with respect to the \index{density!matrix}density matrix, $\tilde \rho (t,t_0)$, whose evolution is defined 
by 
$i \hbar d \tilde \rho (t,t_0) / d t =  [\hat H _{\mathrm{eff}}(t),\tilde \rho (t,t_0)]$,
where $\hat H _{\mathrm{eff}}(t)$ is an {\it effective} Hamiltonian chosen so as to {\it exactly} generate the equation of motion for the \index{field operator!non-condensate}non-condensate operator of Eq.~(\ref{allen_eqn:noncondeom}).
It can be split as
$\hat H_{\mathrm{eff}} (t) = \hat H_{\rm HF} (t) + \hat H ' (t)$, 
where
$
\hat H _{\rm HF} (t) = \int d {\mathbf{r}} \hat \psi'^\dagger [ -\hbar^2 \nabla^2/2m + U_{\rm{eff}}(\mathbf{r},t)] \hat \psi'$
is the unperturbed \index{mean-field theory! Hartree--Fock}HF Hamiltonian\index{Hamiltonian, expression for!Hartree--Fock};
the perturbation term $\hat H ' (t)$
arises from the difference between contributions of multiples of
\index{field operator!non-condensate}non-condensate operators and their mean values; see
Refs.~\cite{zaremba_nikuni_99,griffin_nikuni_book_09,proukakis_jackson_08} for more details. 
The equation of motion for $f({\bf{p}},{\bf{r}}, t)$ thus takes the form
\begin{equation}
\label{allen_eqn:densevolve2}
\begin{split}
\frac{\partial f ({\bf{p}},{\bf{r}}, t)}{\partial t} =& \frac{1}{i \hbar} {\rm{Tr}}
\left\{ \tilde \rho (t,t_0) \left [\hat f ({\bf{p}},{\bf{r}}, t_0), \hat H_{\mathrm{eff}}(t) \right]\right\},
 \\
 =& \frac{1}{i \hbar} {\rm {Tr}} \left\{ \tilde \rho (t,t_0) \left [\hat f ({\bf{p}},{\bf{r}}, t_0), \hat H_{\rm HF}(t) \right] \right\}
\\ &+ \frac{1}{i \hbar} {\rm {Tr}} \left\{ \tilde \rho (t,t_0) \left [\hat f ({\bf{p}},{\bf{r}}, t_0), \tilde H'(t) \right]\right\}.
\end{split}
\end{equation}
The first term in Eq.~\eqref{allen_eqn:densevolve2} [$\sim O(g)$] generates the usual free-streaming contributions to the \index{quantum! Boltzmann equation}Boltzmann equation in the presence of a slowly varying external
potential $U_{\rm{eff}}({\mathbf{r}},t)$, with a gradient expansion in terms of differentials in position, ${\bf{r}}$ and momentum, ${\bf{p}}$;
the second term [$\sim O(g^2)$] encapsulates the particle evolution due to collisions and gives rise to \index{collisional!integral}collisional integrals.
We thus obtain the 
following kinetic \index{quantum! Boltzmann equation}Quantum Boltzmann Equation (QBE) (see
e.g.\ \cite{zaremba_nikuni_99,jaksch_gardiner_97,luiten_reynolds_96,holland_williams_97})
\begin{equation}
\frac{\partial f}{\partial t} + \frac{\bf{p}}{m} \cdot \nabla_{\bf{r}} f- (\nabla_{\bf{r}}U_{\rm{eff}}) \cdot (\nabla_{\bf{p}}f)= C_{12}[f, \phi] + C_{22}[f];
\label{allen_eqn:qbe}
\end{equation} 
the 
right hand side of Eq.~(\ref{allen_eqn:qbe}) contains {\it two} \index{collisional!integral}collisional integrals representing different types of collisions 
between condensate and \index{non-condensate!atom}non-condensate atoms\index{collision(s)!condensate-thermal particle exchanging}\index{collision(s)!condensate--non-condensate|see{collision(s), condensate-thermal}}\index{collision(s)!thermal-thermal}
\cite{kirkpatrick_dorfman_83,kirkpatrick_dorfman_85a,kirkpatrick_dorfman_85b,kirkpatrick_dorfman_85c};
here thermal excitations are treated \index{semiclassical approach}semiclassically via Eq.\ (\ref{allen_eqn:hfenergies}).

The \index{collisional!integral}collisional integrals involve binary atomic collisions which lead to the scattering of particles from initial to final states, where $f_i \equiv
f({\bf{p}}_i,{\bf{r}}, t) $ is the statistical factor for the destruction of a particle in state
$i$ and $(f_i + 1)$, for the stimulated creation of a particle in state $i$.  
They are defined as follows: 
\begin{multline}
C_{12}[f,\phi] = \frac{4 \pi}{\hbar} g^2|\phi|^2  \int \frac{d {\bf{p_2}}}{(2\pi \hbar )^3}\int
\frac{d {\bf{p_3}}}{(2\pi \hbar )^3}\int \frac{d {\bf{p_4}}}{(2\pi \hbar )^3}
( 2\pi \hbar )^3
 \delta (m {\bf{v}}_{\mathrm{c}} + {\bf{p}}_2 - {\bf{p}}_3 - {\bf{p}}_4)  
 \\ 
 \begin{aligned}
 &\times 
 \delta(\varepsilon_{\mathrm{c}} + \tilde \varepsilon_2 - \tilde \varepsilon_3 -\tilde \varepsilon_4)
(2\pi\hbar)^3[\delta({\bf{p}} - {\bf{p}}_2) - \delta({\bf{p}} - {\bf{p}}_3) -
\delta({\bf{p}} - {\bf{p}}_4)] \\ 
& \times  [(f_2+1)f_3f_4 - f_2(f_3+1)(f_4+1)],
\end{aligned}
\label{allen_eqn:c12}
\end{multline}
refers to a \index{collision(s)!condensate-thermal particle exchanging}collision involving one condensate atom with one non-condensate atom and results in
atom transfer between the two subsystems, and
\begin{equation}
\begin{split}
C_{22}[f] = &\frac{4 \pi}{\hbar} g^2  \int \frac{d {\bf{p_2}}}{(2\pi \hbar )^3}\int \frac{d
{\bf{p_3}}}{(2\pi \hbar )^3}\int \frac{d {\bf{p}}_4}{(2\pi \hbar )^3} 
 (2\pi \hbar )^3 \delta ({\bf{p}} + {\bf{p}}_2 - {\bf{p}}_3 - {\bf{p}}_4) 
\\ & \times  \delta(\varepsilon + \tilde \varepsilon_2 - \tilde \varepsilon_3 -\tilde
\varepsilon_4)
[(f+1)(f_2 + 1)f_3f_4  - ff_2(f_3+1)(f_4+1)]
\end{split}
\label{allen_eqn:c22}
\end{equation}
refers to a \index{collision(s)!thermal-thermal}collision involving the redistribution of two non-condensed atoms.

In the absence of the collisional terms $C_{12}$ and $C_{22}$ in Eq.\
(\ref{allen_eqn:qbe}), the equation is known as the \index{Vlasov equation}Vlasov equation, arising in diverse
fields of physics such as plasma physics, condensed matter physics and astrophysics.
$C_{22}$ is the usual collisional term appearing in the ordinary bosonic \index{Boltzmann equation}Boltzmann equation (in the absence of condensation). Setting $C_{22}[f]=0$, as appropriate in \index{equilibrium! thermodynamic}thermodynamic equilibrium, implies that particles are distributed according to the \index{Bose!--Einstein distribution}Bose--Einstein distribution $f_i \rightarrow n_{\rm BE}(\tilde \varepsilon_i) = [ e^{\beta (\tilde \varepsilon_i - \mu)} - 1 ]^{-1}$.
In the presence of a condensate, $C_{12} = 0$ when the condensed and non-condensate\index{non-condensate!component} subsystems are in local \index{equilibrium!diffusive}`diffusive' equilibrium, implying that there is no {\it net} transfer of particles between the two subsystems on average. In cases
out of equilibrium, this term acts to transfer atoms between the condensed and non-condensed
parts of the system\index{collision(s)!condensate-thermal particle exchanging}, with overall particle \index{number conservation}number conservation ensured by the source
term in Eq.~(\ref{allen_eqn:dgpe}) being given by,
\begin{equation}
R ({\mathbf{r}},t) = \frac{\hbar}{2|\phi ({\mathbf{r}},t)|^2} \int \frac{d {\bf{p}}}{(2\pi \hbar )^3}
C _ {12}[f({\bf{p}}, {\bf{r}},t), \phi({\mathbf{r}},t)].
\label{allen_eqn:source}
\end{equation}

The delta functions of Eqs.~(\ref{allen_eqn:c12}) and~(\ref{allen_eqn:c22}), enforce conservation of
energy and momentum; condensate
atoms appearing in $C_{12}$ have energy  $\varepsilon_{\mathrm{c}} =m v_{\mathrm{c}}^2/2 + \mu_{\mathrm{c}}$, and momentum $m{\mathbf{v}}_{\mathrm{c}}$,  where $\mu_{\mathrm{c}}$ is
the condensate \index{chemical potential}chemical potential.
Although the exact expression for $\mu_{\mathrm{c}}$ depends on pair and \index{anomalous average!triplet}triplet anomalous averages \cite{zaremba_nikuni_99}, within \index{ZNG}ZNG their static values are not included in $\mu_{\mathrm{c}}$ or in the excitation energies (which are both treated to first order in $g$), thus yielding the reduced expression
\begin{equation}
\mu_{\mathrm{c}} (\mathbf{r},t) = -\frac{\hbar^2 \nabla^2 \sqrt{n_{\mathrm{c}}(\mathbf{r},t)}}{2m \sqrt{n_{\mathrm{c}}(\mathbf{r},t)}} + V_{\mathrm{ext}}({\bf{r}}) + gn_{\mathrm{c}} (\mathbf{r},t) + 2g n' (\mathbf{r},t).
\label{allen_eqn:chempot}
\end{equation}
It is crucial to note that this does {\it not} imply that
the simultaneous annihilation or creation of more than one particle are ignored, since such terms are {\it implicitly} maintained (in an approximate, yet consistent manner) to second order in $g$ in the dynamical equations, thus giving rise to the \index{collisional!integral}collisional integral $C_{12}$. While this procedure is approximate, the method used here is internally self-consistent, and eventually leads to a theory which conserves both total energy and {\it total} particle number.

To make contact with physical variables, we note that the \index{non-condensate!density}non-condensate density can be reconstructed from the distribution function via
\begin{equation}
 n' ({\bf{r}}, t) = \int \frac{d {\bf{p}}}{(2\pi \hbar)^3} f({\bf{p}},{\bf{r}}, t).
\label{allen_eqn:fnrelation}
\end{equation}

Equations~(\ref{allen_eqn:dgpe}) and (\ref{allen_eqn:qbe}) 
are the closed set of \index{ZNG}ZNG equations for a condensate coexisting in a trap with a cloud of
thermal excitations.  Their solution governs the 
evolution of 
the physical variables of interest. 

\subsection{Numerical Implementation}
\label{allen_sec:numerics}
Equation~(\ref{allen_eqn:dgpe}) is solved in much the same way as the simpler, $T=0$ \index{Gross--Pitaevskii equation}GPE,
with the \index{non-condensate!mean-field}non-condensate mean field and the \index{source term}source term calculated from the \index{QBE|see{quantum, Boltzmann equation}}\index{quantum! Boltzmann equation}QBE at each time step;
we briefly outline the numerical procedure for solving the \index{quantum! Boltzmann equation}QBE~(\ref{allen_eqn:qbe}) in the mean-field-dominated regime, where the full distribution function $f({\bf{p}},{\bf{r}}, t)$ is required to accurately describe the \index{thermal!cloud}thermal cloud --- see~\cite{jackson_zaremba_02a} for details.

We simulate the evolution of the \index{thermal!cloud}thermal cloud phase space distribution by means of a set of
\index{test particles}test particles.  This allows a Lagrangian approach to be adopted, whereby the motion of a
phase point is followed in time according to Newton's equations of motion $d
\mathbf{r}_i(t)/dt =\mathbf{p}_i(t)/m$ and $\mathbf{p}_i(t)/dt = - \nabla U_{\rm{eff}}
({\mathbf{r}},t)$.
These are initially distributed in phase space according to $f({\bf{p}},{\bf{r}}, t) \simeq \gamma (2 \pi \hbar)^3 \sum_{i=1}^{N _ {tp}} \delta(\mathbf{
r} - \mathbf{r}_i(t)) \delta (\mathbf{p} - \mathbf{p}_i(t)),$
where $\gamma = N'/N _ {tp}$ is a weighting factor which ensures that the physical number
$N'$ of thermal atoms is being represented.  As \index{test particles}test particles are only used in order to accurately sample the distribution, there is no need for their number to match the actual number of thermal atoms, which can be smaller or larger, with the only requirement that
$N_{tp}$ needs to be sufficiently large to minimise the effects of using a discrete particle description. 

The density at a particular grid point can be determined by using a cloud-in-cell
method~\cite{hockney_eastwood_book_88}.  This method takes into account the actual position of a
particle within a cell, and also allows for migrations between grid cells.  As residual \index{fluctuations!thermal}fluctuations
remain, additional smoothing is necessary, achieved through convolving the densities with a Gaussian of width a few times the \index{thermal!cloud}thermal cloud grid spacing (the condensate density is also convolved for consistency except in simulations involving \index{soliton!dark}solitons and \index{vortex}vortices, where the main physical effects would become smeared out).
At the end of
each time step, the probability that a given test particle suffers a collision is determined and its position and velocity are updated accordingly (respecting momentum and \index{energy conservation}energy conservation).  The collisions are treated using a \index{Monte Carlo}\index{quantum!Monte Carlo}Monte Carlo sampling
technique: whether a \index{collision(s)}collision has occurred, and what its type is, is determined by comparison of the calculated collisional probability with a random number.
Finally, the condensate evolution is appropriately amended by the occurrence of any collisions which transfer particles into/out of the condensate. 
As collisions are dealt with separately from the dynamical evolution, we are able to choose which collisional effects to include in our simulations, and thus assess the importance of each contribution in a physical setting, thereby enabling a comparison to other theories (in appropriate limits).

\subsection{\index{ZNG}ZNG Condensate \index{hydrodynamic(s)!equation}Hydrodynamics}

Introducing phase and amplitude variables\index{field operator!density-phase representation} to the \index{condensate!wavefunction}condensate wavefunction $\phi (\mathbf{r},t) = \sqrt{n_{\mathrm{c}}(\mathbf{r},t)} e^{i \theta (\mathbf{r},t)}$
allows Eq.\ (\ref{allen_eqn:condeom}) to be recast in \index{hydrodynamic(s)!equation}hydrodynamic form: 
\begin{equation}
\frac{\partial n_{\mathrm{c}}}{\partial t} = \nabla \cdot (n_{\mathrm{c}} {\mathbf{v}}_{\mathrm{c}}) = - \Gamma_{12}[f,\phi]
\hspace{0.4cm} \mbox{and} \hspace{0.4cm}
m \frac{\partial {\mathbf{v}_{\mathrm{c}}}}{\partial t} = - \nabla \left(\mu_{\mathrm{c}} + \frac{1}{2} m v_{\mathrm{c}}^2 \right) \;.
\label{allen_eqn:hydro2}
\end{equation}
The condensate velocity is given by 
${\bf{v}}_{\mathrm{c}}(\mathbf{r},t)\equiv (\hbar/m) \nabla \theta (\mathbf{r},t)$,
$\mu_{\mathrm{c}}$ is defined by Eq.\ (\ref{allen_eqn:chempot})
and  $\Gamma_{12}[f,\phi] 
\equiv \Gamma_{12} ({\mathbf{r}},t) = 2 n_{\mathrm{c}} R(\mathbf{r},t) / \hbar$ is the `source' term due to particle-exchanging collisions between condensate and \index{thermal!cloud}thermal cloud.

\section{Validity Issues \label{allen_validity}} 
\subsection{Validity Domain}
A key feature of a successful theory for describing bosonic quantum fluids is its ability to
explain the phenomenon of superfluidity. This arises naturally within \index{ZNG}ZNG, since the condensate velocity $v_{\mathrm{c}}$ is precisely the velocity describing the flow of the \index{superfluid!density}superfluid density in the \index{two-fluid hydrodynamic model}two-fluid model. Although the \index{ZNG}ZNG equations are based on the \index{Beliaev}Beliaev scenario, this is done here within certain approximations for the excitations, namely that they are single-particle-like (\index{mean-field theory! Hartree--Fock}Hartree--Fock) and they ignore \index{anomalous average!pair}\index{anomalous average!triplet}anomalous averages.  Thus, the present implementation is
inappropriate for describing collective \index{phonon}phonon-like excitations which become important at very low
temperatures. This could nonetheless be remedied, in favour of \index{Bogoliubov!quasiparticle}Bogoliubov \index{quasiparticle!Bogoliubov|see{Bogoliubov, quasiparticle}}quasiparticles, since
the nature of the \index{condensate!excitation}excitation spectrum is intrinsic to the choice of the Hamiltonian
on which perturbation theory is applied; 
hence the formalism presented here could also be systematically extended within the \index{Beliaev}Beliaev field-theoretic approach to account for such features (see Appendix A of Ref.~\cite{zaremba_nikuni_99}, Chapters 3 and 17 of Ref.~\cite{griffin_nikuni_book_09} and Ref.~\cite{proukakis_jackson_08}).

The \index{ZNG}ZNG theory has been remarkably successful in describing a range of dynamical problems, across an extremely broad temperature range, in both the \index{collisionless regime}collisionless and \index{hydrodynamic(s)!regime}hydrodynamic regimes; it even describes the rather involved  problem of condensate growth \cite{bijlsma_zaremba_00}, {\it provided} an initial `seed' condensate is {\it assumed}. However, its present implementation will fail in the regime of critical \index{fluctuations!thermal}fluctuations, i.e.\ very close to the transition temperature  
in 3d systems, or over broader temperature ranges in 1d/2d systems which exhibit enhanced \index{fluctuations!phase}phase \index{fluctuations!thermal}fluctuations.

\subsection{Relevance to Other Theories}
The \index{ZNG}ZNG theory is formally related to
the approach of Walser {\emph{et al.}}~\cite{walser_williams_99,walser_cooper_00}, K{\"o}hler {\emph{et al}.}~\cite{kohler_burnett_02} and Proukakis {\emph{et
al.}}~\cite{proukakis_burnett_96,proukakis_burnett_98,proukakis_01}, which amounts to a similar perturbative treatment starting however from an appropriately generalised basis which includes the \index{anomalous average!pair}pair anomalous average. However, an important advantage of \index{ZNG}ZNG is its formulation in terms of phase space distribution functions, which allows for a relatively straightforward numerical implementation, that can be extended to the \index{hydrodynamic(s)!regime}hydrodynamic regime. 
(The closely-related explicitly \index{number-conserving}number-conserving approach~\cite{gardiner_97,gardiner_morgan_07,castin_dum_98} has not yet been formulated in terms of a self-consistently evolving \index{thermal!cloud}thermal cloud.)

This book also describes a number of \index{stochastic!method}stochastic classical-field approaches (see
e.g.~\cite{stoof_99,blakie_bradley_08,lobo_sinatra_04,brewczyk_gajda_07}).
In some sense, the \index{ZNG}ZNG theory can be thought of as the analogue of the full theory of Stoof~\cite{stoof_99}, when derived within a \index{symmetry-breaking}symmetry-breaking perspective.
The key difference of the \index{ZNG}ZNG formulation compared to those of \index{classical field}classical field methods is that the \index{ZNG}ZNG approach directly describes the {\it condensate} \index{order parameter}order parameter, whereas the field appearing in the \index{Gross--Pitaevskii equation}GPE-like equation of \index{classical field}classical field theories describes a {\it range} of `classical' or `coherent' modes which include, but are {\it not restricted to}, the condensate mode.  In such theories, the latter is to be extracted (in the Penrose--Onsager sense\index{Penrose--Onsager!definition of condensation}) by numerical diagonalisation {\it a posteriori}\index{Penrose--Onsager!condensate mode! extraction in c-field methods}.
In contrast to this,  the \index{ZNG}ZNG method {\it directly} generates two distinct components, the condensate and
the non-condensate\index{non-condensate!component} (see Fig.\ \ref{allen_fig2}), and thus it provides an intuitive picture of the fundamental physical processes taking place in a partially-condensed system; this also guarantees  a direct link to the superfluid properties of the system --- which are harder to extract from unified \index{stochastic!method}stochastic approaches --- thus leading directly to established theories of \index{superfluid!helium}superfluid helium (see Section~\ref{allen_relevance}).
On the implementation front, the current importance of \index{ZNG}ZNG is that it includes a fully dynamical \index{thermal!cloud}thermal cloud self-consistently\index{collision(s)!thermal-thermal}; this is crucial, e.g.\ in order to correctly predict the Kohn mode induced when suddenly displacing the trap, whereas \index{stochastic!method}stochastic methods describing a band of modes coupled to a {\it static} \index{thermal!cloud}thermal cloud would actually lead to artificial \index{damping}damping.

\begin{figure}[b]
\centerline{\includegraphics[scale=1]{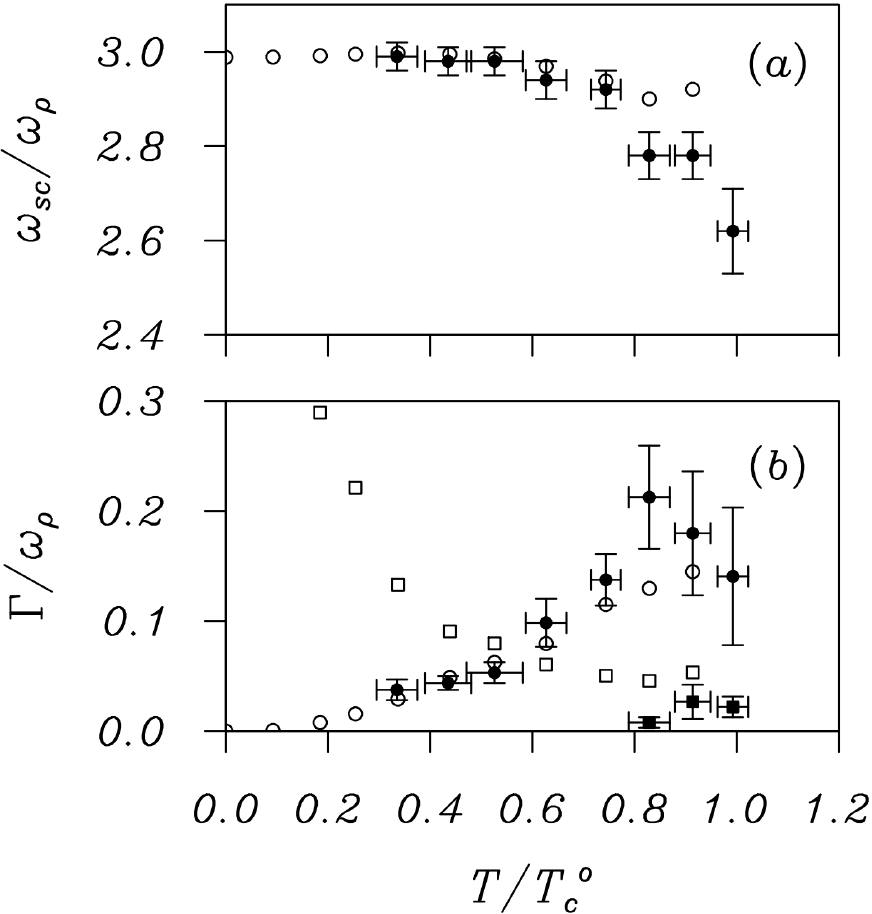}}
\caption{ (a) Frequency and (b)  damping rate of the scissors mode for a variable total number of atoms, intended to
simulate the experiment of Ref.~\cite{marago_hechenblaikner_01}.  The condensate mode is indicated by open (theory) and filled
(experiment) circles.  The open squares in (b) show the calculated average damping rate of the two thermal cloud
modes, while the filled squares are the corresponding experimental values.  
Reprinted with permission from
B. Jackson and E. Zaremba, {\em Finite-temperature simulations of the scissors mode in
Bose--Einstein condensed gases}, Phys. Rev. Lett. {\bf 87}, 100404 (2001) \cite{jackson_zaremba_01}.
Copyright (2001) by the American Physical Society.
}
\label{allen_fig1}
\end{figure}

\section{Applications \label{allen_applications}}
\subsection{Damping of Condensate Scissors Mode}
\label{allen_sec:app1}
By analogy with earlier work in nuclear physics (see~\cite{iudice_97} for a review), the superfluid nature of a BEC can be probed by observing 
the so-called `\index{scissors mode}scissors mode'; this can be excited~\cite{marago_hopkins_00} by adiabatically rotating an axially symmetric trapping potential of the form $V_{\mathrm{ext}}({\mathbf{r}}) = m(\omega^2_{\perp}
\rho^2 + \omega_z^2 z^2)/2$, through a small angle $\theta_0$ about the $y$-axis, and then suddenly in the opposite direction
through an angle $-2\theta_0$.  The observation of a single frequency signals the irrotational flow of the condensate.  This should be contrasted to the superposition of two frequencies
for the thermal component, whose flow has both rotational and irrotational character. 
Simulations with the \index{ZNG}ZNG theory 
by Jackson and Zaremba~\cite{jackson_zaremba_01} found
excellent agreement for $T < 0.8T_{\mathrm{C}}^0$ with a subsequent experiment~\cite{marago_hechenblaikner_01} measuring the temperature dependence of the frequency and \index{damping}damping
of such oscillations. 
The inclusion of collisional processes associated with both $C_{12}$ and $C_{22}$ 
is crucial for explaining the experimental observations, and simulations performed without
collisions result in up to $50\%$ lower \index{damping}\index{damping!rate}damping rates.  
The \index{Landau!damping}Landau \index{damping}damping process is thus a product of both \index{collision(s)}collisional and mean-field effects.

\begin{figure}[b]
\centerline{\begin{tabular}{cc}
\includegraphics[scale=0.9,clip]{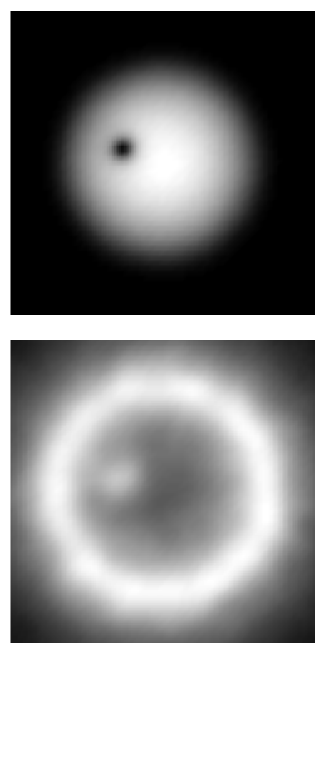} &
\includegraphics[scale=0.375,clip,angle=90,angle=90,angle=90]{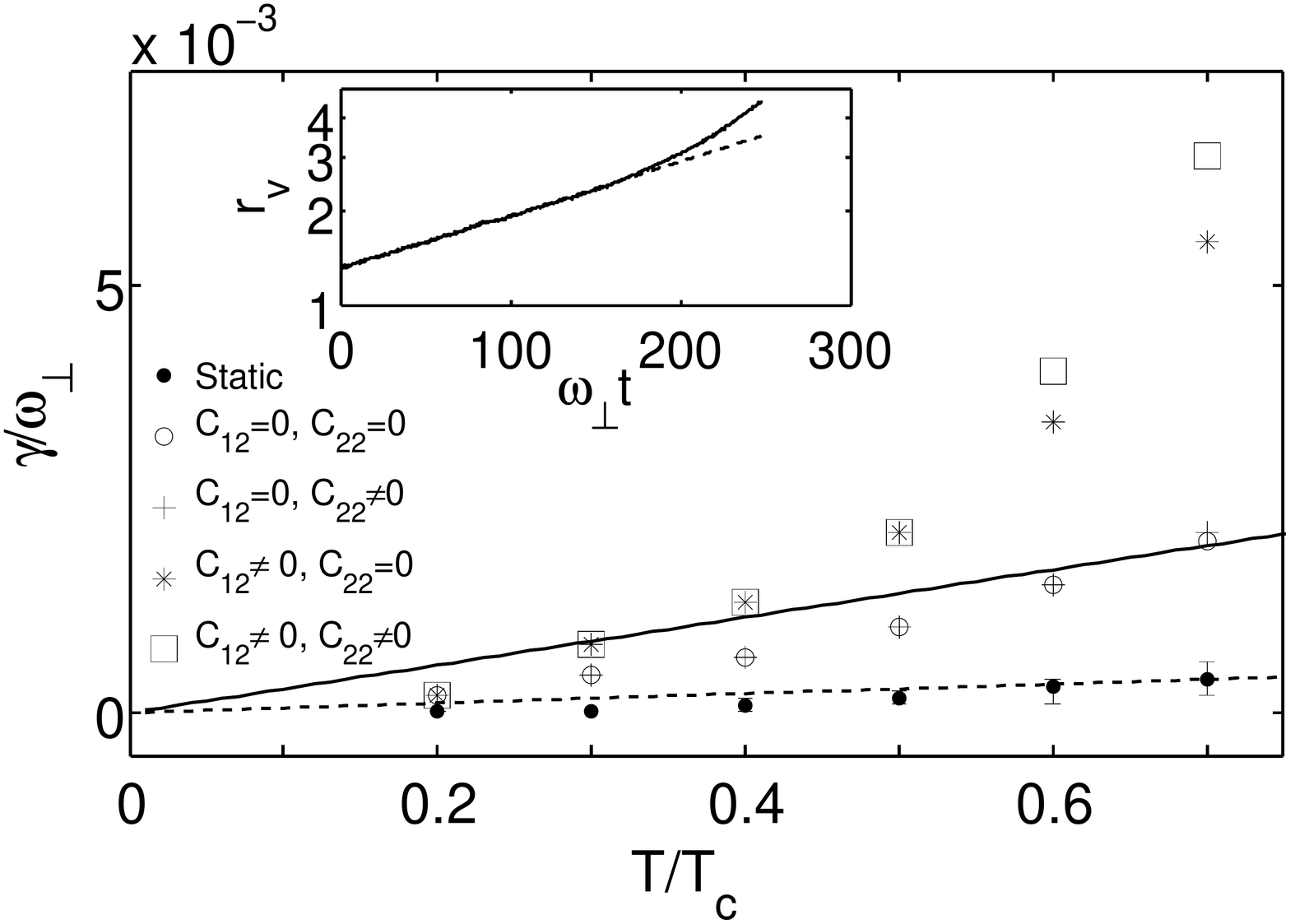}
\end{tabular}}
\caption{Left:  Density cross sections of the condensate (top) and thermal cloud (bottom) 
($T=0.5T_{\mathrm{C}}$) showing clearly the two distinct components simulated within \index{ZNG}ZNG.  Colours range from black (low density) to white (high density) with different scales for the two images. 
     Right:  (Inset) Log-linear plot of vortex radial position as a function of time for $T=0.7T_{\mathrm{C}}$ (solid line); the dashed line is an exponential fit, $r_v(t) = r_0e^{\gamma t}$, to the data over $0 \le \omega_{\perp}t \le 50$.  
(Main Plot)
Values of $\gamma$ based on different levels of approximation: static thermal cloud with condensate dissipation, $iR$, included (solid circles), thermal cloud allowed to evolve within the \index{quantum! Boltzmann equation}QBE (Eq.\ (\ref{allen_eqn:qbe})) without collisions (open circles), with only thermal-thermal \index{collision(s)!thermal-thermal}collisions (pluses), with only particle-transferring \index{collision(s)!condensate-thermal particle exchanging}collisions (stars) or with all collisional processes (squares).
For
comparison, analytic predictions of FS \cite{fedichev_shlyapnikov_99} (solid
line) and DLS \cite{duine_leurs_04} (dashed line) are shown.
(Calculations performed at fixed condensate number, $N_{\mathrm{c}} \simeq 10,000$, and an initial radial vortex offset $r_0 \simeq
0.26R_{\mathrm{TF}}$ from the trap centre).}
\label{allen_fig2}
\end{figure}

\subsection{Decaying Vortex Dynamics}
\label{allen_sec:app2} 
The \index{thermal!cloud}thermal cloud leads to dissipation of an off-centred \index{vortex!finite temperature dynamics}vortex at finite temperatures, causing it to minimise
its energy in a harmonically-trapped condensate by moving out of the condensate radially~\cite{rosenbusch_bretin_02,abo-shaeer_raman_02}.  
Figure~\ref{allen_fig2} (right) shows the nonlinear increase of the decay rate with increasing temperature, clearly
highlighting the role of all collisional processes, particularly at higher temperatures~\cite{jackson_proukakis_09}.
The largest increase in
\index{damping}damping rates arises from the  
particle-exchanging $C_{12}$ \index{collision(s)!condensate-thermal particle exchanging}collisions.
Our results are compared to the analytic predictions of Fedichev and Shlyapnikov
\cite{fedichev_shlyapnikov_99} (FS), which account for the scattering of thermal excitations from the \index{vortex!finite temperature dynamics}vortex core, and those of Duine, Leurs, and Stoof \cite{duine_leurs_04} (DLS), which include the effects of $C_{12}$ collisions within the static \index{thermal!cloud}thermal cloud approximation; the latter are in full agreement with our results, in the corresponding limit
of condensate dissipation from a static \index{thermal!cloud}thermal cloud. The enhanced decay predicted by the full theory highlights the importance of including all dynamical
processes in modelling experiments. 
(See also Ref.~\cite{jackson_proukakis_07} for the corresponding dark soliton analysis accurately reproducing the experiment \cite{burger_bongs_99}.)

\section{Relevance to Other Systems \label{allen_relevance}}
By construction, based on its \index{symmetry-breaking}symmetry-breaking formulation,
the \index{ZNG}ZNG approach is consistent with prevailing theories for the description of \index{superfluid!helium}superfluid helium (see also Ref.~\cite{griffin_nikuni_book_09}).
In the \index{hydrodynamic(s)!regime}hydrodynamic (or collision-dominated) regime,
when the \index{thermal!cloud}thermal cloud enters a state of local \index{hydrodynamic(s)!equilibrium}hydrodynamic equilibrium,
\index{Landau}Landau's famous 2-fluid equations for the \index{superfluid!density}superfluid and normal densities arise as a special case of the \index{ZNG}ZNG equations \cite{zaremba_nikuni_99,nikuni_zaremba_99,griffin_nikuni_book_09},
when the condensate and \index{thermal!cloud}thermal cloud are in diffusive local equilibrium with the same \index{chemical potential}chemical potential.
These equations can be extended to include \index{hydrodynamic(s)!damping}hydrodynamic \index{damping}damping, which arises when the distribution functions deviates from its local equilibrium form, giving rise to the 
\index{Landau!--Khalatnikov}Landau--Khalatnikov \index{two-fluid hydrodynamic model}two-fluid
hydrodynamic\index{hydrodynamic(s)!equation} equations~\cite{nikuni_griffin_01a},
of \index{liquid helium}liquid helium, with the correct transport coefficients, thus facilitating a precise determination of the crossover between the mean-field \index{collisionless regime}collisionless and \index{two-fluid hydrodynamic model}two-fluid hydrodynamic regime.  
It would be of interest to extend the \index{ZNG}ZNG formalism to other situations such as \index{dipolar}dipolar Bose gases, \index{spinor condensate}spinor condensates, bosonic \index{mixture!Bose--Bose}mixtures and apply this scheme to trapped superfluid \index{Fermi!gas}Fermi gases.

\section*{Acknowledgments}
Eugene Zaremba gratefully acknowledges Tetsuro Nikuni and Allan Griffin for the development of the theory and Brian Jackson and Anssi Collins for its numerical implementation.  Nick Proukakis additionally acknowledges related discussions with Keith Burnett and visiting Professorships at Queen's University, Kingston and University of Toronto.  
This work was funded by the UK EPSRC (AJA, CFB, and NPP) and NSERC of Canada (EZ).


\end{document}